\AtBeginDocument{%
  \paperwidth=\dimexpr
    1in + \oddsidemargin
    + \textwidth
    + 1in + \oddsidemargin
  \relax
  \paperheight=\dimexpr
    1in + \topmargin
    + \headheight + \headsep
    + \textheight
    + 1in + \topmargin
  \relax
  \usepackage[pass]{geometry}\relax
}

\RequirePackage{fix-cm}
\documentclass[twocolumn,epjc3]{svjour3} 
\usepackage{graphicx}
\usepackage{amsmath}
\usepackage{float}
\usepackage{tensor}
\usepackage{amssymb}
\usepackage{widetext}
\usepackage{subcaption}
\usepackage{color}
\begin{document}

\title{Non-comoving Cold Dark Matter in a $\Lambda$CDM background
}


\author{Sebasti\'an N\'ajera         \and
        Roberto A. Sussman 
}


\institute{Sebasti\'an N\'ajera \at
              Instituto de Ciencias Nucleares, Universidad Nacional Aut\'onoma de M\'exico
(ICN--UNAM),\\ A. P. 70 –- 543, 04510 M\'exico D. F., M\'exico. \\
              \email{sebastian.najera@correo.nucleares.unam.com}           
}
\journalname{Eur. Phys. J. C}
\date{Received: date / Accepted: date}

\maketitle

\begin{abstract}

We examine the evolution of peculiar velocities of cold dark matter (CDM) in localized arrays of inhomogeneous cosmic structures in a $\Lambda$CDM background that can be identified as a frame comoving with the Cosmic Microwave (CMB). These arrays are constructed by smoothly matching to this cosmological background regions of Szekeres-II models whose source is an imperfect fluid reinterpreted as non-comoving dust, keeping only first order terms in $v/c$. Considering a single Szekeres-II region matched along two comoving interfaces to a $\Lambda$CDM background, the magnitudes of peculiar velocities within this region are compatible with values reported in the literature, while the present day Hubble expansion scalar differs from that of the $\Lambda$CDM background value by a 10\% factor, a result that might provide useful information to the ongoing debate on the $H_0$ tension. While the models cannot describe the virialization process, we show through a representative example that structures of galactic cluster mass reach the onset of this process at redshifts around $z\sim 3$.
 
\end{abstract}

\section{Introduction}
\label{intro}

It is a well known fact that cosmic structures at different scales are not comoving with a frame of reference associated with the CMB and identified as the frame of a $\Lambda$CDM background. This fact follows from measured and inferred peculiar velocities between CDM structures and the CMB \cite{Ellispec}. These peculiar velocities are clearly non--relativistic (up to 3000 \hbox{km/s}), which justifies studying their dynamical evolution by means of Newtonian gravity as a good approximation \cite{Elst}. However, relativistic effects might not be negligible when considering  the superposition of  non-relativistic peculiar velocities on scales comparable with the Hubble horizon. Studies of large-scale peculiar velocities \cite{Strauss,Kashl} have shown a connection between them and the anisotropy and inhomogeneity on large but still subhorizon scales. 

Since the scale dependence of these velocities is still an open topic, it is worth studying this dependence by means of a general relativistic approach that complies with their non-relativistic magnitudes. This approach involves considering congruences of observers with different 4--velocities, leading to distinct energy momentum tensors so that peculiar velocities result from  momentum and energy fluxes between the congruences \cite{Ellis}. Different congruences of observers define different Hubble flows characterized by the kinematic quantities associated with their 4-velocities. In particular,  Tsagas has examined the effect of considering peculiar velocities under this approach on the interpretation of cosmological observations (see \cite{Tsagas,Tsagas2} and references therein). Another example of non--comoving matter can be found in \cite{gaspar2019}, which examined the evolution of cosmic voids formed by baryons and CDM. 

The simplest example of an observed effect that can be attributed to non--comoving observers is the peculiar velocity of our local Hubble flow with respect to the CMB frame, which manifests itself in the large observed CMB temperature dipole. In particular, we can define the CMB frame as the one in which this dipole vanishes \cite{Bolejkoart}. Besides relativistic models of peculiar non-relativistic velocities in the context of non--comoving observers \cite{Elst}, there are non-perturbative models based on exact and numerical solutions of Einstein's equations that consider non--comoving observers in complete generality (tilted models), \cite{Tilted2,gaspar2019,Tilted1}.

Since spherical symmetry is too idealized and limited, it is useful to consider the class of exact solutions derived by Szekeres \cite{Plebanski,Krasinski} that (in general) do not admit isometries and thus enhance the available degrees of freedom for applications in cosmology. These solutions are classified in two classes (I and II, Szekeres--I and Szekeres--II hereafter), each one of which subdividing in three subclasses: quasi--spherical, quasi--flat and quasi--hyperbolic, depending on their symmetrical limits.  

The quasi--spherical Szekeres--I models are regarded as the most suitable for cosmological applications and thus have been widely used to address the limitations of the spherically symmetric Lema\^\i tre--Tolman--Bondi (LTB) models (see comprehensive discussion in \cite{Plebanski,Krasinski,bolejkobook}) In these  models mass--energy and all physical and geometric objects appear as the superposition of a dipole on top of the LTB monopole of spherical symmetry \cite{Plebanski}, thus allowing for the description of two or more structures: typically a central monopole (over--density or void) evolving together with an elongated wall type structure (``pancakes'') corresponding to the dipole. This extra degrees of freedom  provided a significant enhancement to the ``Big Void'' models that were proposed ten years ago to account for observations without resorting to dark energy or a cosmological constant \cite{Sussbol}. More recently, these models were used to describe multipole configuration involving an arbitrary number of structures in a $ \Lambda $CDM background \cite{Sussdel,Susshid}, providing an appealing coarse grained rendering of CDM structures at 100 Mpc scales that can be made consistent with the observed cosmography  \cite{Susshiddel}. 

While most applications to cosmology involve the quasi--spherical Szekeres--I models (see review in \cite{bolejkobook}), Szekeres--II models have also been used for studying inflationary scenarios \cite{Barrow} and probing the structure growth factor \cite{Ishak1,Ishak2}. However, we believe that the potential of Szekeres--II models for cosmological applications has remained largely unexplored.

In a recent paper \cite{Matching} we proved that quasi--plane Szekeres-II models admit a smooth matching, along an arbitrary number of matching interfaces, with spatially flat FLRW models. This fact leads to appealing toy models of evolving arrays of multiple inhomogeneous and anisotropic ``pancake--like'' cosmic structures (regions of Szekeres--II models) embedded in a homogenous and isotropic background (we briefly review these models in section \ref{sec:pancake}). 

In the present paper we extend the results of  \cite{Matching} by considering Szekeres-II regions whose energy--momentum tensor has an imperfect fluid form with nonzero energy flux, thus generalizing the exact ``heat conducting dust'' solutions found by Goode \cite{Goode}, whose source is no longer interpreted as a dissipative fluid (difficult to justify for CDM sources), but as dust in a non--comoving frame (neglecting the subdominant baryon contribution) with non-relativistic peculiar velocities. Since the resulting Szekeres--II regions can be matched to a $\Lambda$CDM background with peculiar velocities vanishing at the matching interfaces, the comoving frame associated with this background can be regarded as the CMB frame in which the dynamical contribution of the photon gas is neglected. A similar model was derived in \cite{Buchert} but only considering Szekeres--II models with a comoving dust source. 

Having set up the models, we find that their free parameters allow for the description of evolving CDM structures falling into the comoving CMB frame with peculiar velocities consistent with observed values. Specific numerical examples are provided in full. By computing the Hubble scalar we show that its contrast with respect to its $\Lambda$CDM background value $H_0$ is entirely determined by the shear tensor, producing fluctuations of $H_0$ of the same order of magnitude $\sim 10\,\%$ difference that has emerged in the ``$H_0$ tension'', though peculiar velocities have a negligible $\sim 0.01$ \,\% effect on present day values of $H_0$ between comoving and non-comoving frames. 

While we show that the parameters allow for an evolution of Szekeres--II structures that is free from shell crossing singularities, we argue that these shell crossings mark the limit of validity of the dust description of CDM which necessarily breaks down at the onset of the virialization process. Hence, we present a numerical example of a CDM structure with mass $M\sim 10^{15}\,\hbox{M}_\odot$ (roughly the mass of a galactic cluster) arriving to the onset of virialization at cosmic times $z\sim 3$ that are compatible with structure formation scenarios derived from numerical n-body simulations \cite{nbody}.      

The section by section description of the paper is as follows. In section \ref{sec:SzekII} we introduce Szekeres--II models and summarize their kinematic and geometric properties. In section \ref{sec:Goode} we present a generalization of the exact solution found by Goode \cite{Goode}, whose energy-momentum tensor was originally interpreted as ``heat conducting dust'', showing in section \ref{sec:Tmunu} that this source can be re-interpreted as non-comoving dust endowed with non-relativistic peculiar velocities ({\it i.e.} $v/c\ll 1$). In section \ref{sec:dynvar} we present the physical, kinematical and geometric variables in dimensionless form and expressed as covariant 'exact' perturbations of a $\Lambda$CDM background. We specify and define the parameters that will be used in  section \ref{sec:results} to examine in full detail the peculiar velocity field and the relevant physical, kinematic and geometric variables of the models, including the evaluation of the contrast in the Hubble scalar due to the inhomogeneity of the models. The range of validity of the models ({\it i.e.} of the dust description for CDM) is examined in section \ref{sec:strucform} providing a numerical example of galactic cluster structures starting to virialize at the expected redshifts. Finally, we provide four appendices: Appendix \ref{sec:Impf} presents the general energy-momentum tensor associated with peculiar velocities between two general fluid congruences with distinct 4-velocities. Appendix \ref{sec:compatibility} discusses the consistency conditions involved in assuming peculiar velocities up to first order in $v/c$. Appendix \ref{sec:junctions} provides a summary of junction conditions between the Szekeres-II regions and the $\Lambda$CDM background (which were derived in \cite{Matching}). Appendix \ref{sec:FLRW} presents the $\Lambda$CDM limit in the parameter space without performing a smooth matching.

\section{Szekeres models of class II: geometric and kinematic properties}\label{sec:SzekII}


Szekeres models of class II (Szekeres-II hereafter) are exact solutions of Einstein's equations characterized by the line element \cite{Matching} (see also \cite{Plebanski,Krasinski})
\begin{eqnarray}
ds^2&=&-dt^2+S^2(t)\left[X^2 dw^2+\frac{dx^2+dy^2}{f^2}\right],\nonumber\\
f&=&1+\frac{k\,[x^2+y^2]}{4},\qquad k=0,\pm 1\label{Szmetric} 
\end{eqnarray}
where  $X=X(t,x^i)$ with $x^i=w,x,y$. The canonical orthonormal tetrad $e_{(\alpha)}^a$ associated to this solution is:
\begin{eqnarray}
e^a_{(0)}=\delta^a_0, \quad e^a_{(w)}=\frac{1}{SX}\delta^a_w,\quad e^a_{(x)}=\frac{f}{S}\delta^a_x,\quad e^a_{(y)}=\frac{f}{S}\delta^a_y.\label{tetrad}\nonumber
\end{eqnarray}
where $g_{ab}\,e_{(\alpha)}^a e_{(\beta)}^b =\eta_{(\alpha)(\beta)}$. In general the models do not admit isometries, hence  all invariant quantities depend on the four coordinates $x^a=t,x^i$. 

The rest frames orthogonal to the comoving 4--velocity $u^a=e^a_{(0)}$ are conformally flat, while the 2--surfaces marked by $t$ and $w$ constant in \eqref{Szmetric} have constant curvature whose sign is given by $k$, leading to three general sub-classes: quasi-spherical ($k=1$), quasi-plane ($k=0$) and quasi-hyperbolic ($k=-1$) models. All sub-classes contain axially symmetric limits, as well as higher symmetry particular cases with spherical ($k=1$), plane ($k=0$) and hyperbolical ($k=-1$) symmetries whose natural homogeneous limits are the spherical, plane and hyperbolic Kantowski--Sachs spacetimes ($X=X(t)$). All models admit smooth matchings with their Kantowski--Sachs sub-cases, though as proven in \cite{Matching}, quasi-plane models admit also a smooth matching with spatially flat FLRW models along 3-dimensional hypersurfaces $w=\textrm{const}$. 

The only nonzero kinematic parameters associated with $u^a=e^a_{(0)}$ are the expansion scalar $\Theta=u^a\,_{;a}$ and shear tensor $\sigma_{ab}=u_{(a;b)}-(\Theta/3)h_{ab}$ given by
\begin{eqnarray}
 \Theta=\frac{\dot{X}}{ X}+\frac{3\dot{S}}{S},\qquad
 \tensor{\sigma}{^a_b} = \Sigma \,\tensor{\xi}{^a_b},\qquad \Sigma=-\frac{\dot{X}}{3X},\label{eq:eqThSig}
\end{eqnarray}
where $\dot X=u^a X_{,a}=X_{,t}$ and $\xi^a_b = \delta^a_w\delta^w_b -3h^a_b = \textrm{diag}[0, -2, 1, 1]$. The expansion tensor $\Theta^a_b$ and its three eigenvalues $\Theta^a_b = \lambda_{(i)}\delta^a_b$ 
\begin{eqnarray} \Theta_{ab}&=&\tensor{h}{_a^c}\tensor{h}{_b^d}\,u_{c;d} = \frac{\Theta}{3}\,h_{ab}+\sigma_{ab},\label{eq:exptens}\\
\lambda_{(1)}&=&\Theta^w_w =\frac{\dot S}{S}+\frac{\dot X}{X},\quad \lambda_{(2)}=\lambda_{(3)}=\Theta^x_x=\Theta^y_y =\frac{\dot S}{S}.\nonumber
 \end{eqnarray}  
provide a covariant description of the kinematic anisotropy by identifying two equivalent principal directions  $\lambda_{(2)}=\lambda_{(3)}$ along $e^a_{(x)}$ and $e^a_{(y)}$, which are clearly different from $\lambda_{(1)}$ along $e^a_{(w)}$. 

The inhomogeneity and anisotropy of the models can also be appreciated from the local rate of change of redshift $z$ along null geodesics \cite{Ellis} 
\begin{equation}\label{eq:eqz}
\frac{d z}{z}=\left[\frac{1}{3}\Theta+\sigma_{ab}k^ak^b\right]\,d\vartheta,
\end{equation}
an expression that must be integrated along light rays, parametrized by the affine parameter $\vartheta$, with tangent null vectors $k^a$ defined by \eqref{Szmetric}. Note that the redshift distribution measured by local observations along a comoving worldline ($x^i$ fixed) is isotropic only if $h_a^b\,\Theta_{,b}=0$ and $\sigma_{ab}=0$ along the worldline.

\section{An exact solution with dust and energy flux}\label{sec:Goode} 

In a comoving frame the models described by \eqref{Szmetric} are compatible with the most general energy-momentum tensor \cite{Matching}
\begin{equation}\label{eq:Tab}
T_{ab} = (\rho + \Lambda) u_a u_b +  (p-\Lambda) h_{ab}+ \pi_{ab}+ 2 q_{(a} u_{b)}, 
\end{equation}
where $\rho,\,p,\,\pi^{ab}$ and $q_a$ are the matter-energy density, isotropic and anistropic pressure and energy flux.  However, we will consider  as source of \eqref{Szmetric} the particular case of \eqref{eq:Tab} with $p=\pi_{ab}=0$ that generalizes to $\Lambda>0$ the exact solution found by Goode \cite{Goode}   
\begin{equation}\label{eq:Tabgoode}
T_{ab}= (\rho +\Lambda) u_a u_b -\Lambda h_{ab} + 2 q_{(a} u_{b)},
\end{equation}
Also, we will consider henceforth only the quasi-plane subcase $k=0$ of \eqref{Szmetric} 
\begin{eqnarray}\label{eq:Szekmet}
ds^2=-dt^2+S^2(t)[X^2 dw^2 + dr^2+r^2d\phi^2],
\end{eqnarray}
where $r,\,\phi$ are cylindrical coordinates defined as $x=r\cos \phi$, $y=r\sin\phi$ (we remark that the quasi-plane subcase is not spatially flat, see \cite{Matching}). The dust density $\rho$ and energy flux vector $q_a = q_r\delta_a^r+ q_\phi\delta_a^\phi$ are given by
\begin{eqnarray}
\kappa(\rho +\Lambda) &=& \frac{3\dot S^2}{S^2}-\frac{X_{,rr}+rX_{,r}-r^2X_{,\phi\phi}}{r^2S^2X}+\frac{2\dot S\dot X}{SX},\label{eq:rho00}\\
\kappa q_r &=& \frac{\dot X_{,r}}{S^2 X},\qquad q_\phi = \frac{\dot X_{,\phi}}{r^2S^2 X},\label{eq:q00}
\end{eqnarray}
with the metric function $X$ given by
\begin{eqnarray}
X&=& A(r,\phi,w)+B(r,\phi ,w)Q(t,w)+F(t,w),\label{eq:eqX}\\
Q&=&c_1(w)+c_0(w)\int \frac{dt}{S^3},\label{eq:eqQdef}\\
A&=&\alpha_3(w)r^2+\alpha_2(w)r\cos\phi +\alpha_1(w) r\sin\phi+\alpha_0(w),\nonumber\\
\label{eq:eqA}\\
B&=&\beta_3(w)r^2+\beta_2(w)r\cos\phi +\beta_1(w)r\sin\phi+\beta_0(w)\nonumber\label{eq:eqB}.\\
\end{eqnarray}
where we notice that $X$ has the form of a monopole with two independent superposed dipoles $A(w,r,\phi)$ and $B(w,r,\phi)$, becoming only a single dipole $A(w,r,\phi)$ in the perfect fluid subcase ($q_a=0$ and thus $Q=0$). 

The main advantage of cylindrical coordinates is dealing with a bounded coordinate $0\leq\phi\leq 2\pi$ and the fact that they mark in a simple way the class of privileged observers along  the curve $r=0$ ($t,\,w$ constant) parametrized by $w$ (or $x=y=0$) that is a geodesic and an integral curve of a Killing vector of the rest frames. Axial symmetry follows by the restriction $\alpha_2=\alpha_1=\beta_1 =\beta_2=0$ in (\ref{eq:eqA})--(\ref{eq:eqB}) so that $X$ becomes independent of $\phi$.

The functions $S(t)$ and $F(t,w)$ are found by solving the differential equations
\begin{eqnarray}
\frac{2\ddot{S}}{S}+\left(\frac{\dot{S}}{S}\right)^2&=&\Lambda,\label{eq:Spun} \\
\ddot{F}+3\left(\frac{\dot{S}}{S}\right)\dot{F}&=&\frac{2}{S^2}[Q\beta_3+\alpha_3]\label{eq:Fddot}.
\end{eqnarray}
with (\ref{eq:Spun}) formally identical to the Friedman equation for the pressure in FLRW  models, thus suggesting an identification of $S(t)$ with an FLRW factor, an identification that we discuss rigorously in section \ref{sec:pancake} and in Appendix \ref{sec:junctions} in terms of a smooth matching of Szekeres-II models with a spatially flat FLRW spacetime along surfaces of constant $w$ (see also \cite{Matching}).

\section{Non-comoving CDM}\label{sec:Tmunu}

Since a thermally dissipative ``heat conducting dust'' source is not appropriate for a late time cosmological model, we propose a wholly different interpretation for \eqref{eq:Tabgoode} as non--comoving cold dark matter (CDM) in which $q^a$  is no longer a heat conducting vector, but an energy flux proportional to the peculiar velocity of a dust source $\hat T^{ab} =(\hat\rho+\Lambda)\,\hat u^a \hat u^b$ where the 4-velocity $\hat u^a$ is related to the comoving 4-velocity $u^a=e^a_{(0)}$ by the generalized boost  
\begin{equation}\label{eq:decom2}
\hat{u}^a=\gamma(u^a+v^a),\quad \gamma=\frac{1}{\sqrt{1-v_av^a}},\quad v_a u^a=0,
\end{equation}
where $v^a$ becomes the peculiar velocity field of CDM associated with $\hat u^a$ with respect to the CMB frame of a $\Lambda$CDM background associated with $u^a$. From \eqref{eq:rhogen}--\eqref{eq:pigen} in Appendix \ref{sec:Impf} this non--comoving dust energy--momentum tensor referred to $u^a$ takes the form \eqref{eq:Tab} with
\begin{eqnarray} 
\rho &=& \gamma^2 \hat \rho+\Lambda,\qquad p = -\Lambda+\frac13\,v^av_a\gamma^2\hat\rho,\label{noncomdust1}\\
q^a &=&\gamma^2\,\hat\rho v^a,\qquad \pi^{ab}=\gamma^2\rho\,v^{\langle a}v^{b\rangle},\label{noncomdust2} 
\end{eqnarray}
with $\rho,\,q^a$ linear and $p,\,\pi^{ab}$ quadratic in $v^a$. However, observed and inferred peculiar velocities of large scale structures with respect to the CMB frame are clearly non-relativistic, thus it is well justified to keep only terms that are first order in $v^a$, so that $\gamma\sim 1+O(v^2),\,p\sim O(v^2)$ and $\pi_{ab}\sim O(v^2)$, leading (up to first order in $v^a$) to the energy--momentum tensor  
\begin{equation}
T^{ab} = (\rho+\Lambda)\,u^a u^b -\Lambda h^{ab}+ 2 \rho \,v^{(a} u^{b)},\label{eq:TabNew}\end{equation}
which coincides with \eqref{eq:Tabgoode} by identifying $q^a \approx \hat\rho\,v^a$ and $\rho \approx \hat\rho$ (since $\gamma \approx 1$). This energy-momentum tensor in the comoving frame can be regarded as an approximation to a more realistic one given by 
\begin{eqnarray} T^{ab} &=& (\rho_{\hbox{\tiny{cmb}}}+\rho+\Lambda)\,u^a u^b +\left(p_{\hbox{\tiny{cmb}}}-\Lambda\right) h^{ab}+ 2\rho \,v^{(a} u^{b)},\nonumber\\ p_{\hbox{\tiny{cmb}}}&=&\frac{\rho_{\hbox{\tiny{cmb}}}}{3},\quad \rho=\rho_{\hbox{\tiny{cdm}}}+\rho_{\hbox{\tiny{b}}}, \end{eqnarray}
where the comoving CMB radiation and non--comoving baryon densities: $\rho_{\hbox{\tiny{cmb}}}$ and $\rho_{\hbox{\tiny{b}}}$ can be neglected in comparison with the non-comoving CDM density and $\Lambda$ ({\it i.e.} $\rho_{\hbox{\tiny{cmb}}}+\rho\approx \rho_{\hbox{\tiny{CDM}}}$). We believe that this interpretation of their energy--momentum tensor furnishes a solid physical and observational connection to the models under consideration. 

To complement the interpretation of energy flux as non--comoving CDM, it is useful to compute the Hubble scalar  for the non--comoving $4$-velocity $\hat\Theta = \hat h^a_b \hat u^b\,_{;a}$ with $\hat u^a$ defined by \eqref{eq:decom2}. In the linear regime of peculiar velocities $v_a v^a/c^2\ll 1$ we obtain the following relation \cite{Tsagas}
\begin{eqnarray}
	\hat \Theta = \Theta +\vartheta,\qquad \vartheta =\hat{h}_a^b v^a\,_{;b},\label{eq:vartheta}
\end{eqnarray}
where $\hat h_{ab}=g_{ab}+\hat{u}_a\hat{u}_b\approx h_{ab}$ and $v^a = q^a/\rho$ (at linear order).   

\section{Dynamical variables}\label{sec:dynvar}

In order to work with dimensionless variables we normalize the dynamical variables with respect to the present day critical density $8\pi G/(3H_0^2c^4)$ where $H_0$ the present day Hubble length (we use geometric units $c=G=1$). The energy density can be expressed as the sum of a purely time dependent $\Lambda$CDM density (a solution of \eqref{eq:Spun}) plus a term depending on all coordinates that can be conceived as an exact fluctuation over this homogeneous background (see comprehensive discussion on this in \cite{Matching})
%
%
\begin{eqnarray}
\Omega^{\rho}&\equiv& \frac{8\pi\rho}{3H_0^2} =\bar\Omega^\rho+\delta^\Omega,\label{eq:denseq}\\
 &{}& \bar\Omega^\rho = \frac{\Omega_0^m}{S^3}+\Omega_0^\Lambda,\qquad  \Omega^\Lambda_{0}=\frac{8\pi\Lambda}{3H_0^2},\label{eq:lamdens}\\
&{}&\delta^\Omega =  \frac{8\pi \left[ X_{,\phi\phi} - rX_{,r}-r^2X_{,rr}\right]}{3H_0^2 r^2 S^2X} - \frac{2 S_{,\tau} X_{,\tau}}{3SX},\nonumber\\
\label{eq:deldens} 
\end{eqnarray}
where a tilde denotes quantities identified with the homogeneous background (see next section) and $S$ is given by the analytic solution of \eqref{eq:Spun} 
\begin{eqnarray}
S(\tau)=\left(\frac{\Omega^m_0}{\Omega^\Lambda_0}\right)^{1/3} \sinh^{\frac{2}{3}}\left(\frac32 \sqrt{\Omega_0^\Lambda} \tau\right),
\label{eq:SLCDM}	
\end{eqnarray}
with $\tau=H_0 t$ the dimensionless time and $\Omega^m_0$ is the Omega factor associated with CDM (neglecting the baryon contribution).  
The peculiar velocity field follows from $q_a=\rho\,v_a$ (first order on $v_a/c$). From the field equations we have $q_a=q_r\delta^r_a+q_\phi \delta^\phi_a$ with $q_w=0$, hence: 
\begin{eqnarray}
\Omega^{q_r}&\equiv \frac{8\pi q_r}{3H_0^2} =\frac{X_{r,\tau}}{3 H_0^2 S^2 X}\quad\Rightarrow\quad v_r =\frac{\Omega^{q_r}}{\Omega^\rho}\label{eq:eqqr}\\
\Omega^{q_\phi}& \equiv \frac{8\pi q_\phi}{3H_0^2} = \frac{X_{\phi ,\tau}}{3 H_0^2 r^2 S^2 X} \quad\Rightarrow\quad v_\phi =\frac{r\Omega^{q_\phi}}{\Omega^\rho}.\label{eq:eqqth}
\end{eqnarray}
To examine the inhomogeneity and anisotropy of Szekeres--II models it is useful to consider the contrasts of the normalized density $\Omega^\rho$ and Hubble scalar with respect to their $\Lambda$CDM values 
\begin{eqnarray}\label{eq:contrast}
\Delta^{\Omega_\rho} &=& \frac{\Omega^\rho(\tau,r,w,\phi)-\bar\Omega^\rho(\tau)}{\bar\Omega^\rho(\tau)}=\frac{\delta^\Omega}{\bar\Omega^\rho(\tau)},\label{contrast1}\\
\Delta^{\cal H} &=& \frac{{\cal H}(\tau,r,w,\phi)-\bar{\cal H}(\tau)}{\bar{\cal H}(\tau)} =\frac{\delta^{\cal H}}{\bar{\cal H}(\tau)},\label{contrast2} 
\end{eqnarray}
where (from \eqref{eq:eqThSig}) 
\begin{eqnarray} {\cal H}=\frac{\Theta}{3H_0},\qquad \bar{\cal H}=\frac{S_{,\tau}}{S}, \qquad \delta^{\cal H}= \frac{X_{,\tau}}{3X},\end{eqnarray}
It is important to remark that (see \cite{Matching}) the quantities $\delta^\Omega$ and $\delta^{\cal H}$ are covariant fluctuations respectively related to the   electric Weyl  and shear tensors ($E_{ab}$ and $\sigma_{ab}$)
\begin{eqnarray} \delta^\Omega=\bar\Omega^\rho\,\Delta^{\Omega_\rho}=-\frac{\xi^{ab}E_{ab}}{3H_0^2},\quad \delta^{\cal H}=\bar{\cal H}\Delta^{\cal H}=-\frac{\xi^{ab}\sigma_{ab}}{6H_0},\nonumber\label{deltas}\\
\end{eqnarray}  
where $\xi^{ab}=e^a_{(w)}e^b_{(w)}-3h^{ab}$. The peculiar velocities are connected to the magnetic Weyl tensor $H^{ab}$
\begin{equation}\eta_{abc}v^c = \frac{2 H_{ab}}{3\Omega_\rho H_0^2},\label{pecvels} \end{equation}
where $\eta_{abc}=-\sqrt{-g}\varepsilon_{abcd}u^d$ is the Levi-Civita antisymmetric volume form. The variables $\delta^{\Omega},\,\delta^{\cal H}$ and $v^a$ determine the inhomogeneity and anisotropy of the Szekeres--II regions through \eqref{deltas}--\eqref{pecvels} in a coordinate independent manner. In fact, these quantities satisfy evolution equations that reduce in the linear limit to covariant dust perturbations in the comoving gauge \cite{Matching}.          

It is important to remark that first order in $v^a$ does not (necessarily) imply that gradients of peculiar velocities are also small. The general conditions for self consistency of the linear approximation to peculiar velocities are presented in Appendix \ref{sec:compatibility}. 
\begin{figure*}
    \centering
        \includegraphics[width=0.35\textwidth]{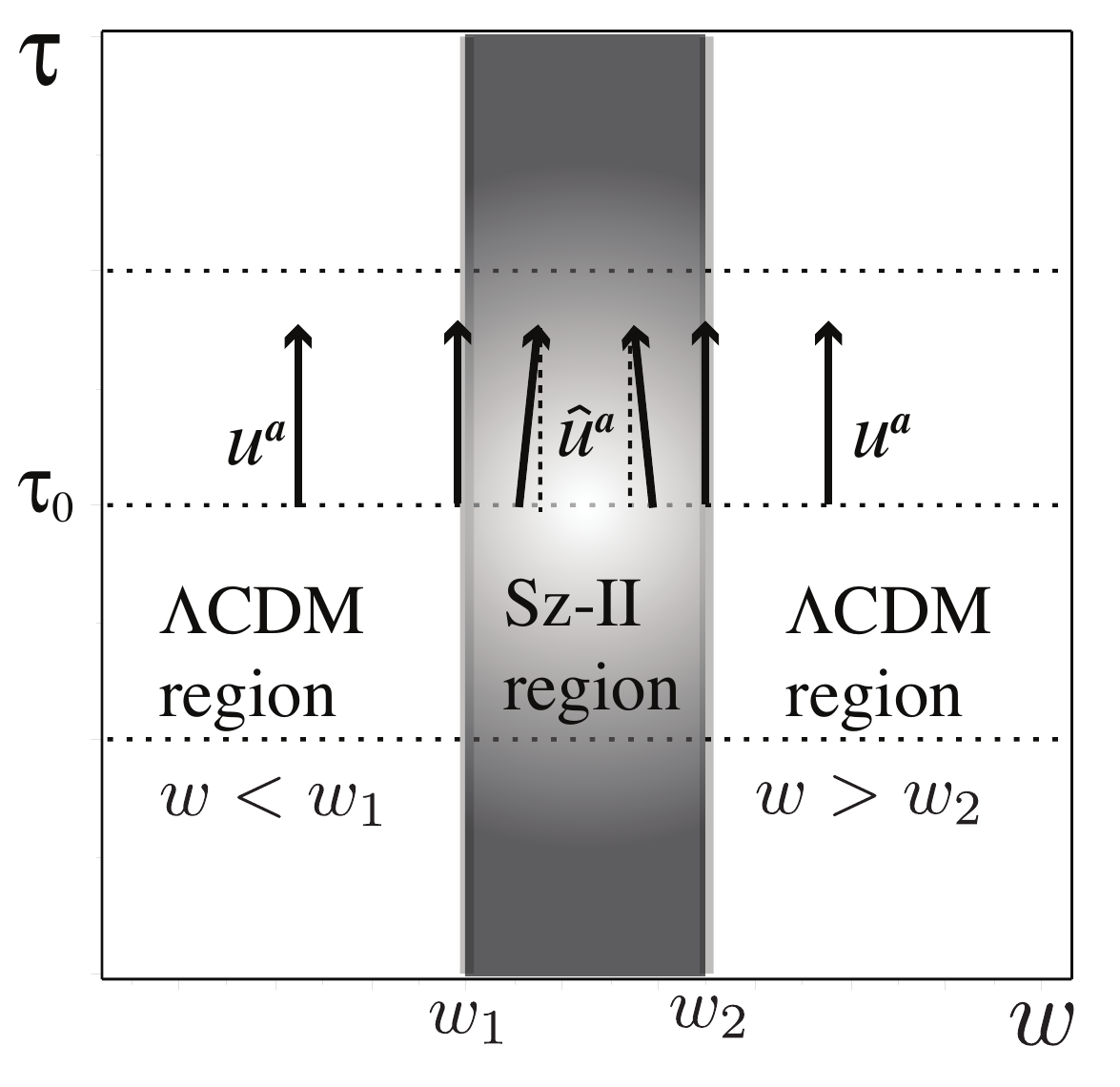}
        \centering
    \caption{Schematic picture of the time evolution of a single Szekeres-II region ``sandwiched'' (smoothly matched)  between two asymptotic $\Lambda$CDM regions (metrics \eqref{eq:Szekmet} and \eqref{eq:FLRW} with coordinates $r,\,\phi$ fixed). The Szekeres-II region extends in the range $w_1<w<w_2$ between the matching hypersurfaces $w_1=0$ and $w_2=2\pi/\nu_0$, with $\nu_0$ defined in the text. Notice how the CDM 4-velocity $\hat u^a$ tilts in the Szekeres-II region with respect to the comoving $u^a$ common to the $\Lambda$CDM regions and the CMB frame.}\label{Fig1}
\end{figure*} 

\section{Smooth matching with $\Lambda$CDM regions}\label{sec:pancake}

\subsection{Pancake models}\label{sec:pancake2} 

Szekeres-II models in general do not satisfy a strict Copernican principle at any scale. However, we can achieve an approximation to a Copernican principle by considering arrays of localized Szekeres--II regions embedded in a spatially flat $\Lambda$CDM background by smooth matchings and thus evolving jointly with it, as in the ``pancake models'' derived and discussed in  \cite{Matching}. In fact, in that paper we presented a brief illustrative example of Szekeres--II regions characterized by an energy-momentum tensor like \eqref{eq:Tabgoode}, with $q^a$ associated with CDM peculiar velocities as in \eqref{eq:TabNew}, but with $\Lambda=0$ (hence the matched FLRW background was an Einstein de Sitter model). We will consider in the following sections the case $\Lambda>0$ of the above mentioned example.  

The regions of a $\Lambda$CDM model to be matched to the Szekeres-II regions are characterized by the following metric (in cylindrical coordinates), energy--momentum tensor and Friedman equation 
\begin{eqnarray}
    ds^2 &=& -dt^2+a^2(t)[dw^2 + dr^2+r^2d\phi^2],\label{eq:FLRW}\\
    T^{ab} &=& (\tilde{\rho}+\Lambda) u^a u^b -\Lambda h^{ab},\\
	\tilde{\Omega^\rho} &\equiv& \frac{8\pi}{3H_0^2}(\tilde \rho + \Lambda)=\frac{\tilde\Omega_0^m}{a^3}+\tilde \Omega_0^\Lambda=\frac{\tilde H^2}{H_0^2},\label{eq:rhoFLRW}\end{eqnarray}
where $\tilde\rho$ is the CDM density, $\tilde H=\dot a/a = \tilde \Theta$ is the Hubble expansion scalar (a tilde denotes FLRW quantities).


As shown in Appendix \ref{sec:junctions} (see also \cite{Matching}), the ``pancake models'' of \cite{Matching} described above rely on the fact that quasi-plane Szekeres-II model with metric \eqref{eq:Szekmet} and the spatially flat $\Lambda$CDM model with metric \eqref{eq:FLRW} admit a smooth matching along an arbitrary number of hypersurfaces ${\cal Z}^i=0,\,\,i=1..n$ parametrized in the cylindrical coordinates of \eqref{eq:FLRW} as $x^a=[t,w_0^i,r,\phi]$ where $w_0^i$ are arbitrary constants. The resulting configurations are sequences of arbitrary numbers of Szekeres-II and $\Lambda$CDM patches separated by matching hypersurfaces $w=w_0^i,\,\,i=1..n$. The junction conditions for these matchings in cylindrical coordinates are (see Appendix \ref{sec:junctions})  
\begin{eqnarray}
S(\tau) &=& a(\tau),\quad X|_{{\cal Z}^i}=1,\nonumber\\
 X_{,\tau}|_{{\cal Z}^i} &=& 0,\quad X_{,r}|_{{\cal Z}^i}= 0, \quad X_{,\phi}|_{{\cal Z}^i}= 0\nonumber\\ 
 X_{,\tau\tau}|_{{\cal Z}^i}&=& 0,\quad X_{,r\tau}|_{{\cal Z}^i}=0,\quad X_{,\phi\tau}|_{{\cal Z}^i}=0, \nonumber\\
X_{,r\phi}|_{{\cal Z}^i}&=& 0,\quad X_{,rr}|_{{\cal Z}^i}=0,\quad X_{,\phi\phi}|_{{\cal Z}^i}=0,\nonumber\\
\label{matchings3}
\end{eqnarray}
where ${}|_{{\cal Z}^i}$ denotes evaluation at $w=w_0^i$ for arbitrary $(\tau,r,\phi)$. These junction conditions must be applied to the functions $F,\,Q,\,A,\,B$ in the analytic form of $X$ given in \eqref{eq:eqX}--\eqref{eq:Fddot}.

It is important to remark that the connection between the Szekeres-II model and a $\Lambda$CDM background that we are considering is based on performing smooth matchings between different Szekeres-II regions with metric \eqref{eq:Szekmet} and $\Lambda$CDM regions of metric \eqref{eq:FLRW} as discussed above (based on \cite{Matching} and illustrated by figure \ref{Fig1}), with the parameters of the Szekeres-II regions only restricted by fulfilling the matching conditions \eqref{matchings3}. This is a completely different approach to reaching an FLRW limit by a sequence of Szekeres-II models whose parameters approach an FLRW spacetime for the full extension of the manifold (as shown in Appendix \ref{sec:FLRW} such limit in the parameter space does exist).

\subsection{Building up a model}\label{sec:themodel}

While the ``pancake model'' configurations described in section \ref{sec:pancake2} allow for multiple Szekeres-II regions that can be different as long as \eqref{matchings3} hold, we will consider the case of a single Szekeres-II region extending along a continuous range of $w$, matched with a $\Lambda$CDM background along timelike hypersurfaces marked by the minimal and maximal values of $w$ in this range. This configuration is depicted schematically in figure \ref{Fig1}. The free parameters of the metric functions \eqref{eq:eqX}--\eqref{eq:eqB} characterizing the Szekeres-II region must be restricted to comply with \eqref{matchings3}. These restrictions are described as follows:      

\begin{description}
\item For the $\alpha$ parameters in \eqref{eq:eqA}\,\, $\alpha_0(w)=\cos^2(\nu_0 w)$, with the remaining free functions given as $\mathcal{C}_i\sin^2(\nu_i w)$, hence the matching interfaces are at $w=w_1=0$ and $w=w_2= 2\pi/\nu_0$, while the arbitrary constants $\mathcal{C}_i$ must comply with $0<\mathcal{C}_i<1$ in order to fulfill the compatibility conditions (see Appendix \ref{sec:compatibility}). We assume $\nu_i\geq 2H_0/c$ to have Szekeres-II sections extending well below the present day Hubble radius, though this length scale can always be modified.    
\item For the $\beta$ parameters in \eqref{eq:eqB} the compatibility conditions and $v_av^a\ll 1$ require $\beta_i^2\ll 1$ and $\beta_i\beta_j\ll 1$ with $i,j=0,1,2,3$, hence these free functions must have the form $\beta_i(w)=\epsilon_i \varphi_i(w)$, with $\varphi_i(w)$ bounded functions and the constants $\epsilon_i$ complying with $\epsilon_i \epsilon_j \ll 1$ and satisfying \eqref{matchings3}. In particular, we consider the choice $\phi_i=\sin^2(\nu_i w)$.
\item In the analytic form (\ref{eq:SLCDM}) of $S(\tau)$ for the $\Lambda$CDM regions we define present cosmic time from $S(\tau_0)=1$ leading to $\tau_0= 0.9662$. We select the values $\Omega^m_0=0.3$ and  $\Omega^\Lambda_0=0.7$ so that present day $\bar\Omega_0=1$. The present day peculiar velocity in \eqref{eq:eqqr}--\eqref{eq:eqqth} and the density and Hubble scalar contrasts in \eqref{contrast1}--\eqref{contrast2} become
\begin{eqnarray}\Delta^{\Omega_\rho}_0 &=& \delta^\Omega_0,\qquad \Delta^{\cal H}_0 = \delta^{\cal H}_0,\label{eq:Deltas0}\\
v_{r0}&=&\left[v_r\right]_0 = \Omega^{q_r}_0,\qquad v_{\phi0}=\left[v_\phi\right]_0 = r\Omega^{q_\phi}_0.\label{eq:vr0}
\end{eqnarray} 
\item The function $F$  follows from the numerical solution of (\ref{eq:Fddot}), with initial conditions given by $F_{,\tau}(\tau_{0},w)=0$ and $F(\tau_{0},w)=\epsilon_{0}\varphi_{0}(w)$, where $\epsilon_{0}$ is a constant that satisfies $0<\epsilon_{0}<1$, while $\varphi_{0}(w)$ is a sinusoidal function complying with the form stated in the specification of the $\beta$ parameters in \eqref{eq:eqB} discussed above. We selected initial conditions at present cosmic time only as a matter of convenience, as they can be chosen at any fixed $\tau$.  Since (\ref{eq:Fddot}) is a second order linear ODE, its general solution must be of the form $F=\varphi_{+}(w)F_{+}(\tau,w)+\varphi_{-}(w)F_{-}(\tau,w)$, guaranteeing that initial conditions $F(\tau,w_i)=F_{,\tau}(\tau,w_i)=0$ can always be fulfilled by fixed values $w=w_i$ at any value of $\tau$.
\end{description} 
It is worth commenting that $X$ has a very weak dependence on the angular coordinate $\phi$, thus it is possible to get a robust notion of all quantities in terms of a single representative angle. After various trials, we choose the following numerical values for constant parameters: $\epsilon_1 = 0.7$, $\epsilon_2, \epsilon_3= \mathcal{C}_{c_1}=\mathcal{C}_{c_0}=0.001$ and $\nu_i=2H_0/c$, while the rest of the constants $\mathcal{C}_i$ were taken as random numbers such that $0<\mathcal{C}_i<1$, with numerical trials showing very weak dependence on the choice of these numbers. In general, it is necessary to test numerically values for the constants $\epsilon_i,\,\mathcal{C}_{c_i}$ and forms or the functions $c_i$ to avoid shell crossings and to obtain peculiar velocities whose magnitudes comply with the range of values for peculiar velocities of large scale structures found in the literature.
%
\begin{figure*}
    \centering
        \includegraphics[width=0.9\textwidth]{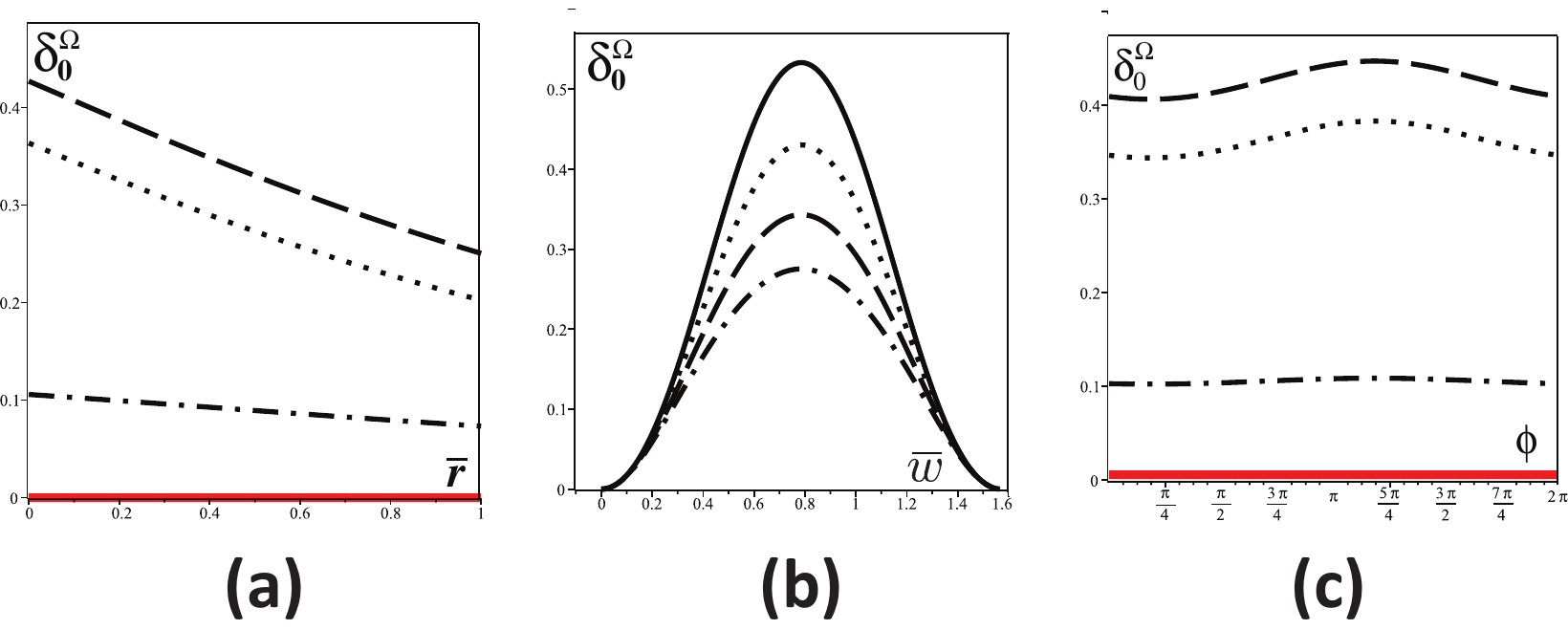}
        \centering
    \caption{Density contrast at present cosmic time for a Szekeres--II region matched to a $\Lambda$CDM background. The panels display $\delta^\Omega_0$ given by \eqref{eq:Deltas0}. Panel (a) as a function of $\bar r=cr/H_0$ and fixed $\phi=\pi/4$, for $w=0, \pi/3, 2\pi/3, 14\pi/15$, respectively depicted by solid, dotted, dashed and dot dashed curves. Panel (b) as a function of $\bar w=cw/H_0$ and fixed $\phi=\pi/4$, for $\bar r=0, 1/3, 2/3, 1$, respectively depicted by solid, dotted, dashed and dot dashed curves. Panel (c) as a function of $\phi$ and fixed $\bar r=0.5$ and fixed values $w=0, \pi/3, 2\pi/3, 14\pi/15$, respectively depicted by solid, dotted, dashed and dot dashed curves.}\label{Fig2}
\end{figure*}
%
%
\begin{figure*}
    \centering
        \includegraphics[width=0.9\textwidth]{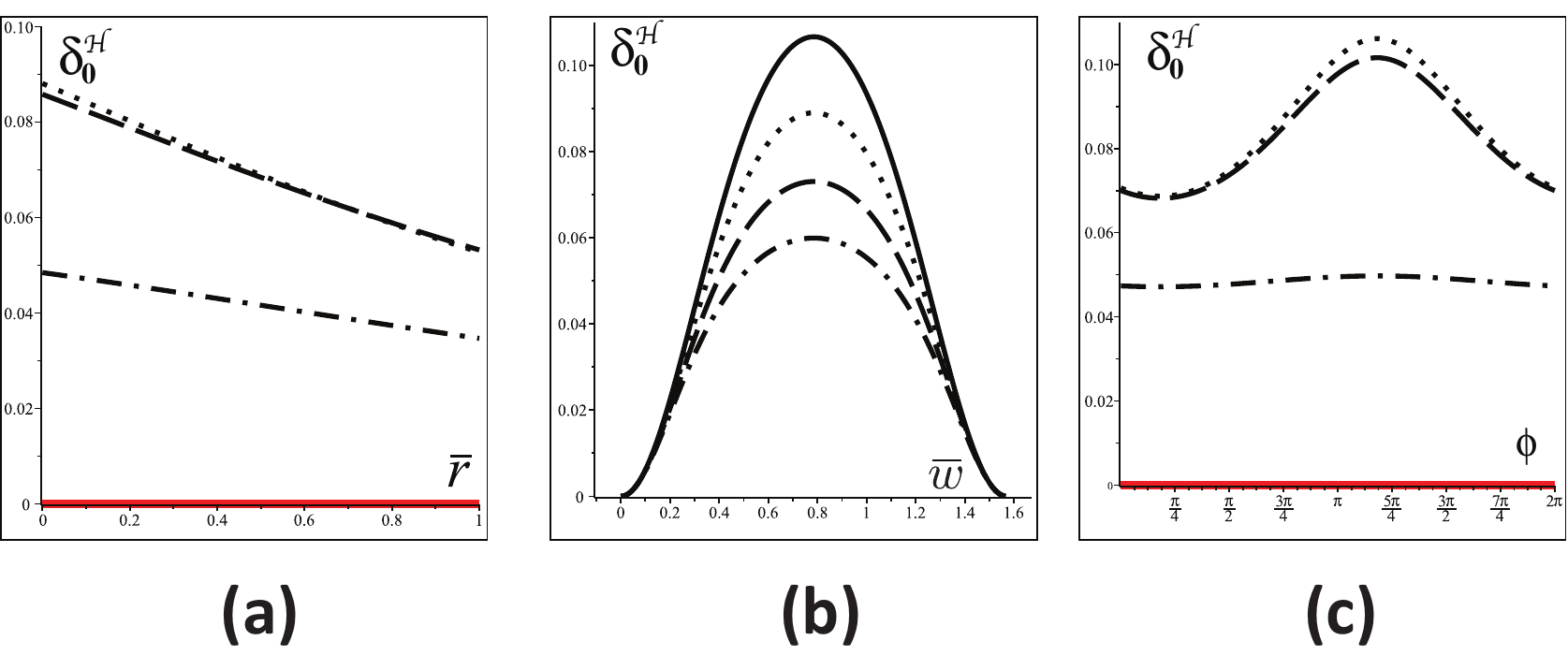}
        \centering
    \caption{Contrasts of the Hubble scalar at present cosmic time for a Szekeres--II region matched to a $\Lambda$CDM background. The panels display $\delta^{\cal H}_0$ given by \eqref{eq:Deltas0}. Panel (a) as a function of $\bar r=cr/H_0$ and fixed $\phi=\pi/4$, for $w=0, \pi/3, 2\pi/3, 14\pi/15$, respectively depicted by solid, dotted, dashed and dot dashed curves. Panel (b) as a function of $\bar w=cw/H_0$ and fixed $\phi=\pi/4$, for $\bar r=0, 1/3, 2/3, 1$, respectively depicted by solid, dotted, dashed and dot dashed curves. Panel (c) as a function of $\phi$ and fixed $\bar r=0.5$ and $w=0, \pi/3, 2\pi/3, 14\pi/15$, respectively depicted by solid, dotted, dashed and dot dashed curves.}\label{Fig3}
\end{figure*}
%
\begin{figure*}
    \centering
        \includegraphics[width=0.7\textwidth]{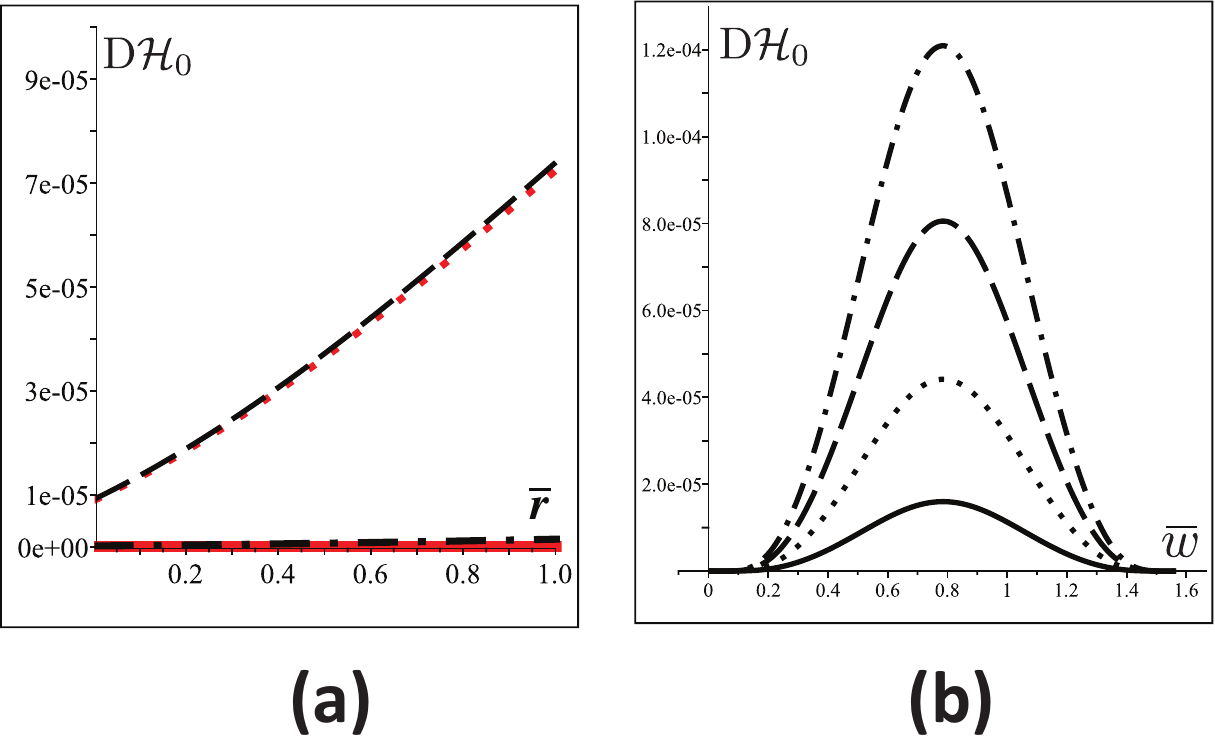}
        \centering
    \caption{Difference between the Hubble scalar in the comoving and non-comoving frames in the Szekeres--II region. The panels display ${\textrm D}{\cal H}_0=\hat{\cal H}_0-{\cal H}_0$ given by \eqref{eq:varhubble} (see also \eqref{eq:vartheta}). Panel (a) as a function of $\bar r=cr/H_0$ and fixed $\phi=\pi/4$, for $w=0, \pi/3, 2\pi/3, 14\pi/15$, respectively depicted by solid, dotted, dashed and dot dashed curves (to improve the depiction  solid and dotted curves are displayed in red). Panel (b) as a function of $\bar w=cw/H_0$ and fixed $\phi=\pi/4$, for $\bar r=0, 1/3, 2/3, 1$, respectively depicted by solid, dotted, dashed and dot dashed curves. Notice in panel (b) that ${\textrm D}{\cal H}_0=0$ as the region matches to the $\Lambda$CDM background}\label{Fig4}
\end{figure*}
%
\begin{figure*}
    \centering
        \includegraphics[width=0.9\textwidth]{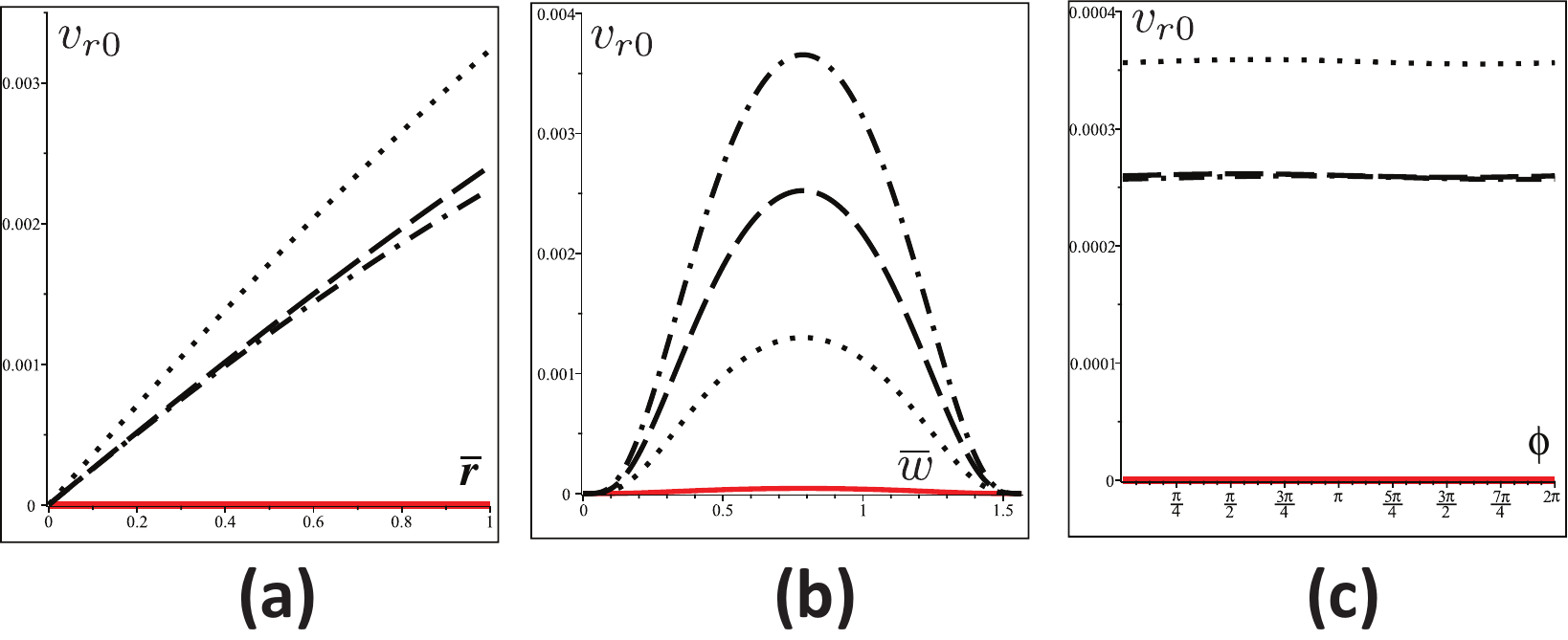}
        \centering
    \caption{Radial peculiar velocities at present cosmic time for a Szekeres--II region matched to a $\Lambda$CDM background. The panels display $v_r$ given by \eqref{eq:eqqr}. Panel (a) as a function of $\bar r=cr/H_0$ and fixed $\phi=\pi/4$, for fixed values $w=0, \pi/3, 2\pi/3, 14\pi/15$, respectively depicted by solid, dotted, dashed and dot dashed curves. Panel (b) as a function of $\bar w=cw/H_0$ and fixed $\phi=\pi/4$, for fixed values $\bar r=0, 1/3, 2/3, 1$, respectively depicted by solid, dotted, dashed and dot dashed curves. Panel (c) as a function of $\phi$ and fixed $\bar r=0.5$ for fixed values $w=0, \pi/3, 2\pi/3, 14\pi/15$, respectively depicted by solid, dotted, dashed and dot dashed curves.}\label{Fig5}
\end{figure*}
%
\begin{figure*}
    \centering
        \includegraphics[width=0.7\textwidth]{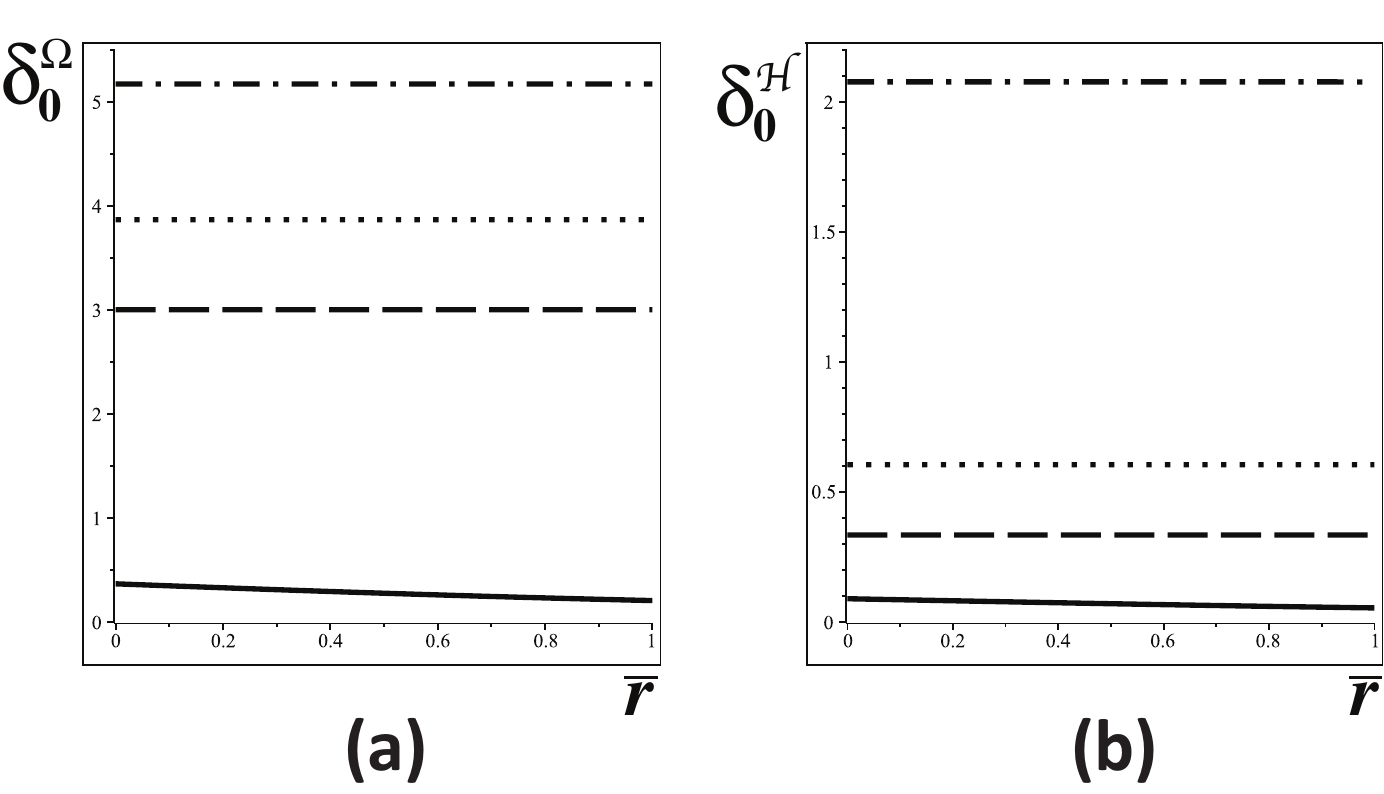}
        \centering
    \caption{Profiles of the density and Hubble scalar contrasts at different cosmic times. The solid, dotted, dashed and dot dashed curve, respectively denote $z=0, 0.5, 1, 2$. All plotted quantities are displayed as functions of $\bar r=cr/H_0$ for fixed values of $w=\pi/3$. Panels (a) and (b) respectively display $\Delta^{\Omega_\rho}$ and $\Delta^{\cal H}$ given by \eqref{contrast1}--\eqref{contrast2}.}\label{Fig6}
\end{figure*}
%
\begin{figure*}
    \centering
        \includegraphics[width=0.7\textwidth]{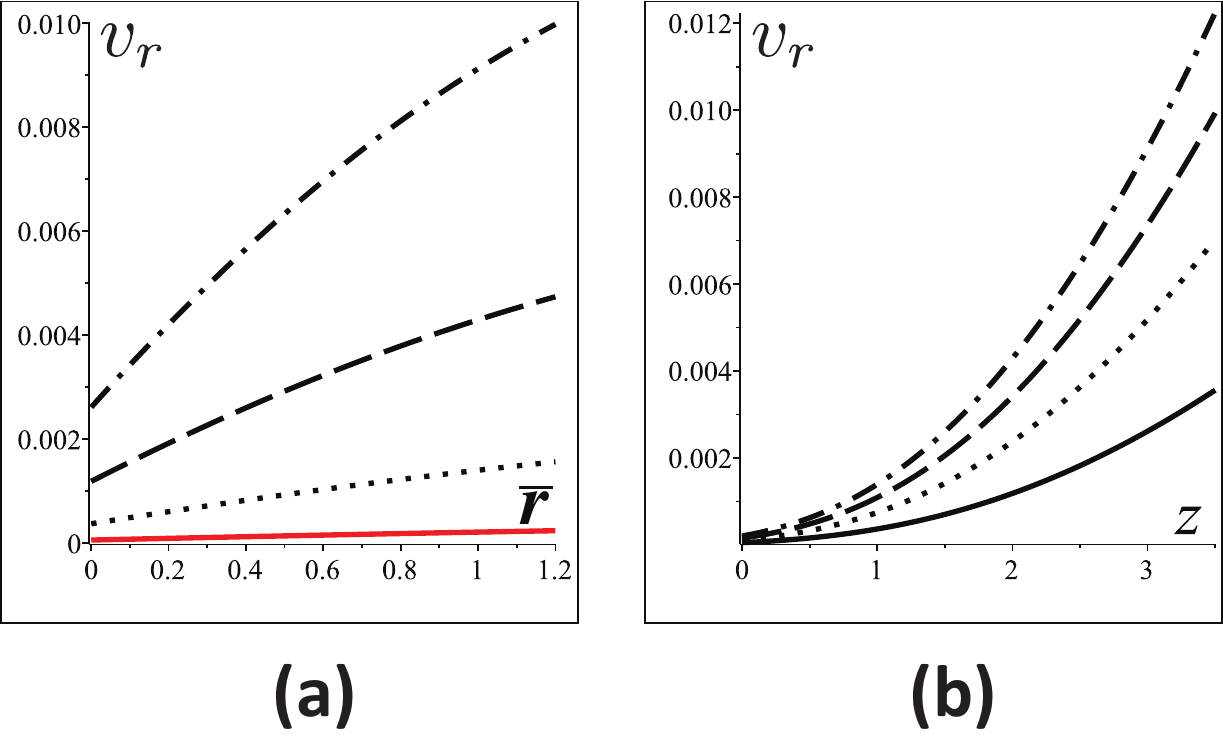}
        \centering
    \caption{Evolution of the radial peculiar velocity $v_r$. Panel (a) displays the profile of $v_r$ for constant $w$ and $z=0, 1, 2 , 3$ (solid, dotted, dashed and dot-dashed curves). Panel (b): evolution of $v_r$ as function of $z$ at $r=0, 1/3, 2/3, 1$ (solid, dotted dashed and dot dashed curves).}\label{Fig7}
\end{figure*}
%
\begin{figure*}
    \centering
        \includegraphics[width=0.4\textwidth]{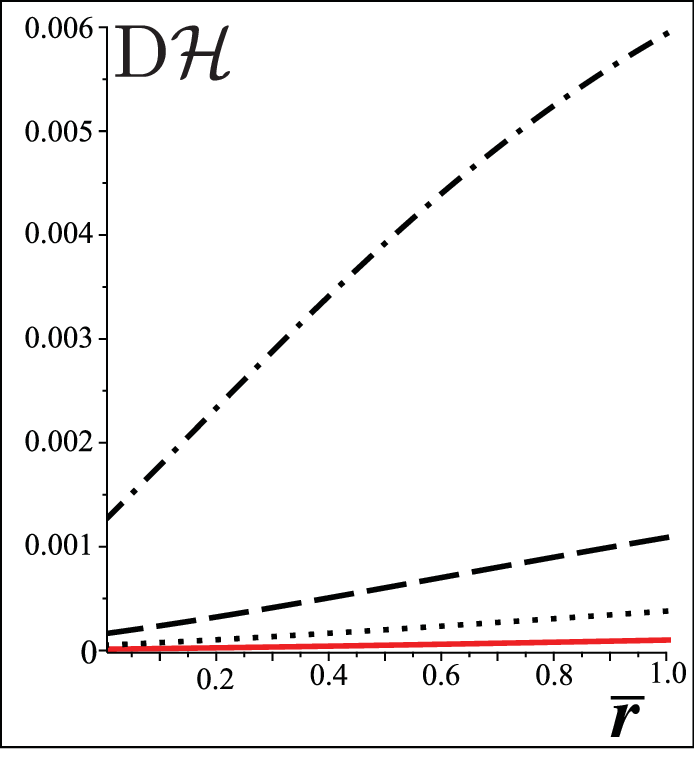}
        \centering
    \caption{The difference between comoving and non--comoving Hubble scalars ${\textrm D}{\cal H}=\hat{\cal H}-{\cal H}$ (see \eqref{eq:varhubble}) at different cosmic times. The solid, dotted, dashed and dot dashed curve, respectively denote $z=0, 0.5, 1, 2$. All plotted quantities are displayed as functions of $\bar r=cr/H_0$ for fixed values of $w=\pi/3$.}\label{Fig8}
\end{figure*}

\section{Results}\label{sec:results}

All plots in figures \ref{Fig2}-\ref{Fig5} correspond to $\tau_0=H_0\,t_0$ fixed at present time $S(\tau_0)=1$. Plotted quantities are displayed as functions of $w$ and $r$ normalized with the Hubble length.  As mentioned before, $w=0$ and $w=\pi$ mark the matching hypersurfaces between the Szekeres-II region and the $\Lambda$CDM background, with the  the Szekeres-II region encompassing the range $0\leq w\leq \pi$ with its edges separated a comoving distance of $\sim 1.5$ times the Hubble radius at $\tau_0$. 

The matching  hypersurfaces depicted in  figure \ref{Fig1} as vertical lines correspond to $w=w_1=0$ and $w=w_2=\pi$ in panels (b) of figures \ref{Fig2}-\ref{Fig5} with the $\Lambda$CDM background extending for $w=w_1<0$ and $w=w_2>\pi$. The horizontal red line depicts the zero contrast level corresponding to the $\Lambda$CDM background.   


Figure \ref{Fig2} displays the profile of the present day density contrast $\delta^\Omega_0$ given by \eqref{eq:Deltas0} as a function of $r$ (panel (a)), $w$ (panel (b)), both normalized by the Hubble length, and $\phi$ (panel (c)). The panels reveal nearly the same value density contrast $\delta^{\Omega}_0\sim 0.4$ with respect to the $\Lambda$CDM background in the directions of $r$ and $w$, with a very weak dependence on $\phi$. While the Szekeres--II regions are clearly inhomogeneous and anisotropic for every observer, these graphs show relatively small local variations of $\delta^\Omega_0$ among observers. In fact, the free parameters allow to adjust the scale variation of the density contrast depending on a desired set of limits. 

Figure \ref{Fig3} depicts the profiles of the present day contrast of the Hubble scalar $\delta^{\cal H}_0$ given by \eqref{eq:Deltas0}, as a function of $r$ (constant $w,\phi$), $w$ (constant $r,\phi$) and $\phi$ (with constant $r,w$). As with the density contrast, the contrast of the Hubble scalar shows small local variation in different directions, as well as weak angular dependence, thus allowing for a a to controlled description of a desired level of inhomogeneity and anisotropy.
Figure \ref{Fig4} displays the difference between present day values of the Hubble scalar in the comoving frame  ${\cal H}_0$ and the non-comoving one $\hat{\cal H}_0$ derived in \eqref{eq:vartheta}
\begin{equation} \textrm{D}{\cal H}_0\equiv \hat{\cal H}_0-{\cal H}_0=\frac{\vartheta_0}{3H_0}=\frac{[\hat h^a_b v^b\,_{;a}]_0}{3H_0},\label{eq:varhubble} \end{equation}
which is valid for $v^a\sim O(v/c)$ so that $\hat h^a_b\approx h^a_b$ up to $O(v/c)$. Panels (a) and (b) respectively display \eqref{eq:varhubble} as a function of $r$ (constant $w,\phi$) and $w$ (constant $r,\phi$) (dependence on the angle $\phi$ is very similar to that displayed in panels (c) of figures \ref{Fig2} and \ref{Fig3}, so it is not displayed). Both panels show that $\textrm{D}{\cal H}_0\sim 10^{-4}$, a value consistent with $v^a\sim O(v/c)$ and with the consistency conditions in Appendix \ref{sec:compatibility}, though it is three orders of magnitude below the 10\,\% associated with the observed $H_0$ tension. 

Figure \ref{Fig5} displays the present day radial velocity $v_{r0}=\left[v_r\right]_0$ given by \eqref{eq:vr0} as a function of $\bar r$ (with $w$ constant), $\bar w$ (with constant $r$) and $\phi$ (with constant $r,w$). Numerical values of the velocities are fractions of $c$, with their magnitude in the  expected range $|v_r| <800$ km/s. As with the contrasts $\delta^{\Omega_\rho}_0$ and $\delta^{\cal H}_0$, the radial velocities have similar values in different directions and very weak angular dependence.

We examine the time variation of the radial profile (for fixed $w$ and $\phi$) of the density and Hubble scalar contrasts \eqref{contrast1}--\eqref{contrast2} in figure \ref{Fig6}. Both of these contrasts take near constant shapes that steadily decrease from $z=2$ to their present values. 
\section{Structure formation}\label{sec:strucform}

Following a careful parameter selection it is possible to obtain configurations free from shell crossings at least up to scales within the Hubble horizon, though some parameter combinations lead to divergent peculiar velocities even without shell crossings. This divergent behavior and the shell crossings signal the limit of validity of the description of CDM as dust. Thus, we restrict the parameters of the models to $|v_r| < 0.01 c$, a reasonable range of validity for structure formation involving non-relativistic conditions, so that spacetime points where this bound is violated can be associated with the onset of virialization whose proper description is beyond the scope of these models.

After several numerical trials we found how to set up the free in order to control the placing the locus marking the beginning of shell crossings (and divergent peculiar velocities) at specific spatial positions and cosmic times measured by redshifts of the $\Lambda$CDM region.  Considering the same choice of parameters as in section \ref{sec:themodel}, we found sufficient parameter freedom to describe structure formation scenarios in which the onset of virialization takes place at redshift  values compatible with observations \cite{Weinberg}, for example, with $|v_r|\to 0.1 c$ at $z\approx 3$. Considering redshifts in the $\Lambda$CDM region given by $S(t)=1/(1+z)$ we plot $v_r$ in figure \ref{Fig7} as function of $r$ and $z$. Notice that the velocities tend to increase their magnitude with increasing $z$. 

We illustrate how peculiar velocities can become larger than the bound $|v_r| < 0.01 c$ for $z<3$ by plotting  the time evolution of $v_r$ in figure \ref{Fig7}. Panel (a) depicts the profile of $v_r$ as function of $r$ ($w$ constant) for $z=0,1,2,3$. The limit velocity $0.01 c$ is reached at $z=3$, marking the onset of virialization. Panel (b) depicts the profile of $v_r$ as function of $z$ with $w$ constantes for various values of $r$. Again, the onset of virialization occurs at $z=3$.   

Finally, we examine in figure \ref{Fig8} the time evolution of the difference between the Hubble scalar in the comoving and non-comoving frames ${\textrm D}{\cal H}= \hat{\cal H}-{\cal H}$ defined in \eqref{eq:varhubble}.  This difference remains small, as expected from the compatibility conditions discussed in Appendix \ref{sec:compatibility}. However, ${\textrm D}{\cal H}$ begins increasing from $z=3$ onwards which sets the limits of validity of the non-relativistic approximation of peculiar velocities relating the two frames. However, we can argue that that times at which the models cease to be valid mark the onset of virialization.  

From the locus of the shell crossing in the example displayed in figure \ref{Fig7} we estimated the approximate conserved mass of the structure undergoing virialization as follows: Considering the energy density from \eqref{eq:rho00} and \eqref{eq:denseq}--\eqref{eq:deldens} for $X$ and $S$ given by \eqref{eq:eqX} and \eqref{eq:SLCDM}, with $F$ and $Q$ obtained by numerical integration of \eqref{eq:eqQdef} and \eqref{eq:Fddot} for the parameters from section \ref{sec:pancake} at the onset of virialization $z=3$, we computed the conserved mass from the following proper volume integral
\begin{equation}\label{eq:eqmass}
	\int \rho(z_i,r,\theta,w) \sqrt{h}\; d^3 x \sim 10^{15}\,\hbox{M}_\odot,
\end{equation}
evaluated at fixed initial redshift $z_i=3$ ({\it i.e.} at a fixed time corresponding to such redshift computed for the $\Lambda$CDM region) and we verified that $v_r\leq 0.1 c$ remained valid along the integration domain. The obtained rest mass roughly corresponds to a galactic cluster whose onset of virialization at $z=3$ is plausible.

\section{Final discussion and conclusions}\label{sec:disc}

We have found for the Szekeres--II models under consideration an appealing physical interpretation as models that describe CDM and dark energy modeled as a $\Lambda$ term with the novelty of incorporating peculiar velocities $v^a=q^a/\rho$ for a non--comoving CDM source, all this in the context of appealing ``pancake models'' of cosmological inhomogeneities described by regions of Szekeres-II solutions embedded by smooth matchings to a $\Lambda$CDM background, introduced in previous work \cite{Matching}. We have also provided a complementary view to previous work looking at the effects of cosmological sources (for example baryons and CDM) evolving along different 4-velocity frames \cite{gaspar2019}).     

In order to illustrate the effects of local inhomogeneity and anisotropy brought by the models we compared their dynamical variables with their values in the $\Lambda$CDM background. For this purpose, we considered a configuration made of a single Szekeres--II region extending 1.5 times the Hubble radius in the $w$ direction,  smoothly matched to a $\Lambda$CDM background on both extremes. We obtained (see figures \ref{Fig2}, \ref{Fig3}) by numerical integration of the field equations the present cosmic time contrasts respect to this background of the density and Hubble scalar ($\delta^{\Omega_\rho}_0$ and $\delta^{\cal H}_0$ from \eqref{eq:Deltas0}), also at different cosmic times (see figure \ref{Fig6}). In all quantities the variation with respect to the angular coordinate was very weak, thus identifying an anisotropy based on differences along (essentially) two directions: $r$ and $w$ (as suggested by looking at the metric \eqref{Szmetric} in rectangular--like coordinates $x,y$ instead of $r,\phi$). 

Figures \ref{Fig2} and \ref{Fig3} reveal present cosmic time values of density and Hubble scalar contrasts within the Szekeres--II region respectively varying from zero to maximal values of $0.4$ and $0.1$ roughly in the same pattern along both directions $r$ and $w$. We tested various combinations of initial conditions and found roughly the same variation patterns with different maximal values, thus indicating a relatively mild deviation of from local isotropy that can be controlled by suitable choices of free parameters at least in scales up  to the Hubble radius. 

Figure \ref{Fig3} reveals that the contrast of the Hubble scalar, $\delta_0^{\cal H}$, exhibits fluctuations with respecto to the background value, $H_0$, of the same order of magnitude $\sim 10\,\%$ associated with the ``$H_0$ tension''. However, peculiar velocities have a negligible effect on $\delta_0^{\cal H}$, which differs by $\sim 0.01$ \,\% when computed for the Szekeres--II model with the same parameters without these velocities ($v^r=v^\phi=0$). This fact stands in agreement with observations indicating that peculiar velocities provide fluctuations of at most $0.1\%$ to the $H_0$ tension, \cite{sedgwick2021effects}. We believe these are interesting results which should be examined in an observational context.

The behavior of present day radial peculiar velocity $v_{r0}=[v_r]_0$ is displayed in figure \ref{Fig5}. Notice how $v_{r0}$ vanishes at $r=0$ and  approximately increases linearly with $r$ for all $w$ within the Szekeres--II region, while in the $w$ direction it takes larger values as $r$ increases (vanishing as required by junction conditions at the $w$ values marking the matching interface). This pattern illustrates how $r=0$ for varying $w$ denotes the coordinate locus of privileged observers, analogous to observers along the symmetry center of spherical symmetry. However, this is not a mere coordinate effect, as the curve along $r=0$ parametrized by $w$ is a spacelike geodesic and Killing vector of the hypersurfaces orthogonal to the comoving 4--velocity (see \cite{Matching}).    
 
We found free parameter choices that lead to an evolution free from shell crossings with peculiar velocities remaining in the non-relativistic regime $|v_r|_0<0.01 c$ at least in scales of the order of the  Hubble radius in the main directions $r$ and $w$. However, we also found free parameter combinations that lead to shell crossings around $z=3$, with peculiar velocities growing and even diverging, thus identifying these spacetime points as marking the onset of virialization when a model based on a dust description of CDM is no longer valid. We estimated a CDM mass of $\sim 10^{15}$ contained in a region associated with these shell crossings that can be identified with a large galactic cluster.

We fully acknowledge the limitations of the models we have studied in this paper: they are basically toy models of inhomogeneities in a $\Lambda$CDM background that are valid only in the scales and cosmic times in which CDM can be modeled as dust. The novelty of our approach (with respect to previous usage of Szekeres models in this context) is that we consider the class Szekeres--II and that CDM is not comoving with the frame associated with the CMB and the $\Lambda$CDM background. Evidently, these toy models cannot describe a highly complex process like virialization, but we can assume that shell crossings (which are a generic feature) can mark the limit of validity of the models due to the onset of this process. 

Nevertheless, we believe that these toy models have a valuable potential for cosmological applications: first, they allow to study the observed non-relativistic peculiar velocities in the framework of an exact solution of General Relativity, thus potentially contributing to improve our understanding of the role of peculiar velocities in cosmic dynamics, not only  at local deep subhorizon scales, but even at scales comparable to the Hubble horizon. Also, a better understanding of the dynamics of peculiar velocities and Hubble flows from different congruences of observers can contribute to address the $H_0$ tension. Finally, the models can serve as an exact solution to probe numerical codes in the emerging field of numerical relativistic cosmology. 

\appendix

\section{Imperfect fluids in terms of peculiar velocities}\label{sec:Impf}

Given a $4$-velocity field the most general form of the energy-momentum tensor is given by \eqref{eq:Tab}
\begin{equation}
T^{ab} = (\rho + \Lambda) u^a u^b +  (p-\Lambda) h^{ab}+ \pi^{ab}+ 2 q^{(a} u^{b)},\nonumber 
\end{equation}
where 
\begin{eqnarray*}
	\rho+\Lambda&=&u_a u_bT^{ab},\quad p-\Lambda =\frac13 h_{ab} T^{ab},\\
	\pi^{ab}=T^{\langle ab\rangle}&=&\left[h^{(a}_ch^{b)}_d-\frac13 h^{ab}h_{cd}\right]\,T^{cd} ,\quad q_a = -u^b T_{ab}
\end{eqnarray*}
with $h^{ab}=u^a u^b+g^{ab}$ being the projection operator, $\rho,\,p,\,\pi^{ab},\,q^a$ are the mass--energy density, the isotropic pressure, the spacelike tracefee anisotropic pressure tensor and the spacelike energy flux vector. 

The energy--momentum tensor (\ref{eq:Tab}) is also referred to as describing ``imperfect fluids'' because of the terms $\pi^{ab},\,q^a$ that are usually identified with dissipative stresses, shear viscosity and heat conduction, in thermal and hydrodynamical systems (see examples in \cite{Krasinski}). However, this interpretation is not suitable for  gravity dominated long range interacting cosmic sources: CDM is best described at large scales as dust, a description also applicable to baryons whose internal energy and thermal dissipative effects are also negligible in these scales \cite{Baryons}. 

A more useful interpretation for the ``imperfect fluid'' terms $\pi^{ab},\,q^a$ in a cosmological context follows as 4--momentum fluxes identified with peculiar velocities associated with a non-comoving 4-velocity with respect to a comoving frame. In particular, we can assume a 4-velocity comoving with a $\Lambda$CDM background for the CMB frame, with CDM and baryons evolving along a 4-velocity field that is not comoving with respect to this frame.

To examine the connection between energy-momentum tensors associated with different frames, we consider two general congruences of spacetime observers with different 4-velocities $u^a$ and $\hat{u}^a$. Choosing $u^a$ as a comoving $4$-velocity, the non-comoving 4-velocity $\hat{u}^a$ is given by the generalization of the Special Relativity boost
\begin{equation}\label{eq:decom}
\hat{u}^a=\gamma(u^a+v^a),\qquad \gamma=\frac{1}{\sqrt{1-v_av^a}}
\end{equation}
where\footnote{Throughout the article we use geometrical units.} $v^a$ is the spacelike peculiar velocity measured by the observer $u^a$ and $v^au_a=0$. Following \cite{Ellisgen} and assuming the energy-momentum for the non-comoving frame to have the general form \eqref{eq:Tab}, the relations between dynamical quantities of the non-comoving and comoving energy-momentum tensors are given by 

\begin{widetext}
\begin{eqnarray}
\rho&=&\hat{\rho}+\Lambda+2\gamma \hat{q}^a v_a+\left\lbrace \gamma^2 v^av_a (\hat{\rho}+\hat{p}) + \hat{\Pi}^{ab}v_a v_b\right\rbrace,\label{eq:rhogen}\\
p&=&\hat{p}-\Lambda+\frac23\,\gamma \hat{q}^a v_a+\frac{1}{3}\left\lbrace \gamma^2 v^av_a (\hat{\rho}+\hat{p}) + \hat{\Pi}^{ab}v_a v_b\right\rbrace\label{eq:pgen}\\
q^a&=&\hat{q}^a+(\hat{\rho}+\hat{p})v^a+\left\lbrace (\gamma-1)\hat{q}^a-\gamma \hat{q}^b v_b\hat{u}^a +\gamma^2 v^b v_b(\hat{\rho}+\hat{p})v^a+\hat{\Pi}^{ab}v_b -\hat{\Pi}^{ab}v_bv_c \hat{u}^a\right\rbrace\label{eq:qgen}\\
\Pi^{ab}&=&\hat{\Pi}^{ab}+\left\lbrace\gamma^2(\hat{\rho}+\hat{p})v^{\langle a}v^{b\rangle}-2 u^{(a}\hat{\Pi}^{b)c}v_c+\hat{\Pi}^{cd}v_cv_d \hat{u}^a\hat{u}^b-\frac{1}{3}\hat{\Pi}^{cd}v_c v_d\hat{h}^{ab}-2\gamma \hat{q}^cv_c u^{(a}v^{b)}+2\gamma v^{\langle a}\hat{q}^{b\rangle}\right\rbrace\label{eq:pigen},
\end{eqnarray}
\end{widetext}
\noindent
where (as in \cite{Ellisgen}) we have written terms linear in $v^a$ outside the curly brackets. Notice that the isotropic and anisotropic pressure are mostly associated with therms that are non-linear (at least quadratic) in $v^a$.

\section{Compatibility conditions}\label{sec:compatibility}

For the sake of completeness we include the compatibility conditions we presented in \cite{Matching}. 

The energy-momentum tensor of a dust source in the frame of a non-comoving observer, $\hat{u}^a$ is
\begin{eqnarray*}
\tensor{\hat{T}}{_a_b}=\rho \hat{u}_{a} \hat{u}_b =\rho\gamma^2 \left( u_a u_b +2 u_{(a}v_{b)}+v_a v_b\right).
\end{eqnarray*}
Using the decomposition (\ref{eq:Tab}), we considered $p=0$, $\Pi_{ab}=0$ and searched under what conditions, in the limit $v_av^a\to 0$, $T_{ab}=\tensor{\hat{T}}{_a_b}$. To order zero, $\lim_{v_av^a} \tensor{\hat{T}}{_a_b} = \rho\left( u_a u_b +2 u_{(a}v_{b)}\right)$, obtaining $\rho u_{(a}v_{b)}=u_{(a}q_{b)}$. Even though we take $v_av^a\ll 1$, this does not imply the derivatives are small, so we must search conditions to first order. As both energy--momentum tensors are conserved, 
\begin{equation}
\tensor{\hat{T}}{^a^b_;_b}-\tensor{T}{^a^b_;_b}=0, \label{eq:derdiff}
\end{equation}
we obtain the second condition when we obtain an identity from this equation $0=0$ in the limit $v^av_a\to 0$. From the previous conditions, and considering the derivatives of the $\gamma$ factor:
\begin{eqnarray}
\tensor{q}{^a_;_b} &=& \tilde\rho_{;b}\gamma^{2}v^a+2\tilde\rho\gamma \gamma_{;b}v^a + \tilde\rho\gamma^{2}\tensor{v}{^a_;_b},
\label{qder}\\
\rho_{;b} &=&\tilde\rho_{;b}\gamma^2+ 2\tilde\rho\gamma \gamma_{;b}.
\label{rhoder}
\end{eqnarray}

From (\ref{eq:derdiff})-(\ref{rhoder}) and the zero order relations it is straightforward to verify that 
\begin{equation}
\tilde\rho\gamma^2(v^a v^b)_{;b}+(\tilde\rho_{;b}\gamma^{2}+2\tilde\rho\gamma\gamma_{;b})v^a v^b=0\label{eq:subs2}.
\end{equation}
Neglecting quadratic terms on $v^a$ we obtain the second condition:
\begin{equation}
(v^a v^b)_{;b}=0.\label{eq:2order}
\end{equation}

This implies $v^a v^b$ is constant, which we take as $v^a v_a <<1$ for consistency with our initial hypothesis $v^a v_a <<1$. Therefore our conditions for compatibility are 
\begin{eqnarray}
\rho u_{(a}v_{b)}&=&u_{(a}q_{b)},\\
(v^a v^b)_{;b}&=&0.
\end{eqnarray}
With these considerations the energy conservation $u_a\tensor{T}{^a^b_;_b}$ is proportional to $(H_0/c)^3$ which justifies our approximation. 

\section{Junction conditions}\label{sec:junctions}

A smooth matching between  two spacetimes, $(\mathcal M_{(+)},g_{(+)})$ and $(\mathcal M_{(-)},g_{(-)})$, which we consider to be  Szekeres-II described by \eqref{Szmetric} and FLRW by  
\begin{equation}ds^2 = -dt^2+\frac{a^2(t)[dw^2 + dx^2+dy^2]}{[1+\frac14\tilde k(w^2+x^2+y^2)]^2},\label{eq:FLRW}\end{equation}
is given by the Darmois conditions \cite{Darmois1,Darmois2} which demand continuity of the first and second fundamental forms at a matching hypersurface ${\cal Z}(x^\alpha)=0$,
\begin{eqnarray}
	[\gamma_{ab}] &= \gamma_{ab}^{(+)}-\gamma_{ab}^{(-)}=0,\quad \gamma_{ab}=g_{ab}+\epsilon n_an_b,\label{eq:Darm1}\\
\left[ K_{a b} \right] &= K_{ab}^{(+)}-K_{ab}^{(-)}=0,\quad K_{ab}=-n_{a;b},\label{eq:Darm2}	 
\end{eqnarray}
where $\gamma_{ab}^{(\pm)} =\gamma\vert_{{\cal Z}_{(\pm)}}=\lim_{w\to w_0^{\pm}}\gamma_{ab}$ (as well as for $K_{ab}$), $n_a$ denotes a unit normal vector to ${\cal Z}$ and $\epsilon =1,-1$ is the vectors tangent to ${\cal Z}$ are (respectively) timelike or spacelike. Considering the identification of coordinates $(t,x^i)$ and orthonormal tetrads in \eqref{Szmetric} and \eqref{eq:FLRW}, we consider the equation $w-w_0=0$, where $w_0$ is an arbitrary constant, to mark the hypersuface given by ${\cal Z}=0$. Then $n_a=e_a^{(a)}=SX\delta_a^w$ and the first fundamental form $\gamma_{ab}$ is parametrized by the coordinates $x^{\alpha}=[t,w_0,x,y]$, where $w$ is fixed to the arbitrary value $w_0$. The first fundamental form has the following components
\begin{eqnarray}
	\gamma_{\alpha \beta}^{(\pm)}&=e_{(\alpha)}^ae_{(\beta)}^b g_{ab}^{(\pm)}, \quad \gamma_{tt}^{(+)}=\gamma_{tt}^{(-)}=-1,\nonumber\\
	\gamma_{xx}^{(+)}&=\gamma_{yy}^{(+)}=\frac{S^2}{f^2},\quad \gamma_{xx}^{(-)}=\gamma_{yy}^{(-)}=-\frac{a^2}{\tilde{f}^2(w_0)}\label{eq:gamcont},
	\end{eqnarray}
	while the non-zero components of the second fundamental form are
	\begin{eqnarray}
	K_{tw}^{(+)}&=&(X\dot S+\dot X S)^{(+)},\quad K_{tx}^{(+)}=(SX_{,x})^{(+)},\nonumber\\
	K_{ty}^{(+)}&=&(SX_{,y})^{(+)},\quad K_{tw}^{(-)}=(\dot a)^{(-)}.\label{eq:Kcont}
\end{eqnarray}
The Darmois conditions combined with \eqref{eq:gamcont}-\eqref{eq:Kcont} imply:
\begin{eqnarray}
	X^{(+)}&=&1,\quad (X_{x})^{(+)}=(X_{y})^{(+)}=(\dot X)^{(+)}=(\ddot X_{x})^{(+)}=0,\nonumber\\
	S(\tau)&=& a(\tau), \quad k=\tilde k=0 \quad \Rightarrow \quad f=\tilde f =1, \label{3junc}
\end{eqnarray}
therefore a smooth matching at $\Sigma$ marked by $w=w_0$ is only possible between quasi-plane Szekeres-II models and a spatially flat FLRW model respectively described by \eqref{eq:Szekmet} and \eqref{eq:FLRW}. The free functions appearing in in $X$ (see \eqref{eq:eqX}--\eqref{eq:Fddot}) must fulfill the matching conditions \eqref{3junc}. Therefore, matching between quasi-flat Szekeres-II and  spatially flat FLRW models can de performed along an arbitrary number of hypersurfaces marked by constant $w$.  Note that these matchings can be performed with a single (but arbitrary) FLRW background, but  the Szekeres--II patches can be different and need not correspond to the same source (as long as conditions \eqref{3junc} hold at the matching hypersurfaces). We refer to such configurations as pancake models (see \cite{Matching} for more detail).  In particular, we considered in this paper the specific pancake configuration in which the FLRW spacetime is a $\Lambda$CDM model matched with a single Szekeres--II region at two hypersurfaces marked by constant $w$. 

\section{The FLRW limit in the parameter space}\label{sec:FLRW}

The form of the line element presented in \cite{Goode} is
\begin{equation}\label{eq:dsGoode}
ds^2=-dt^2+S^2(t)[e^{2\nu}(dx^2+dy^2)+X^2dw^2]
\end{equation}
where $X=e^\nu[A+BQ]+F$, and $e^\nu=(1+k(x^2+y^2))^{-1}$. The perfect fluid case arises for $B=0$, \cite{Goode}, this election can be made considering the form of $B$ given by Goode,
$$B=\beta_3(w)r^2+\beta_2(w)r\cos\theta +\beta_1(w)r\sin\theta+\beta_0,$$
by taking $\beta_3(w)=h_2(w)=\beta_1(w)=\beta_0=0$. Krasinski presents the Szekeres--II with a perfect fluid source with the following line element
\begin{equation}\label{eq:kras}
ds^2=-dt^2+e^{2\alpha}dw^2+e^{2\beta}(dx^2+dy^2),
\end{equation}
where $e^\beta =e^\nu S(t)$, $e^\alpha =\lambda(t,w)+S(t)\Sigma(x,y,w)$, and $\lambda$ a function determined by a differential equation. From the form of the functions stated in \cite{Goode} and \cite{Krasinski} it is easy to show that $\lambda(t,w)=S(t)F(t,w)$, while $\Sigma=Ae^\nu$. Therefore, the condition stated by Krasinki at the end of section 2.1.1 holds. As mentioned in the introduction there is no natural FLRW limit for this model, quoting \cite{Krasinski} ``the
FLRW limit results unnaturally in this subfamily: the additional symmetries of the
FLRW models appear from nowhere.'' Further comments on how the FLRW models arise are found in \ref{sec:FLRW}. It is worth noting that the FLRW limit is a limit in the space of parameters and not a limit in the manifold or an extension of the manifold itself.

\begin{acknowledgements}
SN acknowledges financial support from SEP–-CONACYT postgraduate grants program and RAS acknowledges support from PAPIIT--DGAPA RR107015. We both thank Celia Escamilla for useful and enlightening discussions.

\end{acknowledgements}

\bibliographystyle{unsrt}
\bibliography{NCF}

\end{document}